# Comparing large-scale graphs based on quantum probability theory

Hayoung Choi[a,*], Hosoo Lee[b], Yifei Shen[a], Yuanming Shi[a]

[a]*School of Information Science and Technology, ShanghaiTech University, Pudong district, Shanghai 200031, China*
[b]*Department of Mathematics, Sungkyunkwan University, Suwon 440-746, Republic of Korea*

**Abstract**

In this paper, a new measurement to compare two large-scale graphs based on the theory of quantum probability is proposed. An explicit form for the spectral distribution of the corresponding adjacency matrix of a graph is established. Our proposed distance between two graphs is defined as the distance between the corresponding moment matrices of their spectral distributions. It is shown that the spectral distributions of their adjacency matrices in a vector state includes information not only about their eigenvalues, but also about the corresponding eigenvectors. Moreover, we prove that such distance is graph invariant and sub-structure invariant. Examples with various graphs are given, and distances between graphs with few vertices are checked. Computational results for real large-scale networks show that its accuracy is better than any existing methods and time cost is extensively cheap.

## 1. Introduction

Graph is one of the most common representations of complex data and plays an crucial role in various research areas and many practical applications. Over the past several decades, enormous breakthroughs have been made while many fundamental problems about graphs are remaining to be solved. Comparing graphs is one of the most important problems with a very long history[45]. In practice, the similarity measure of graphs (or equivalently dissimilairty) is widely applied in social science, biology, chemistry, and many other fields. For instance, the similarity measure of graphs can be used to classify ego networks[38], distinguish between neurological disorders[8], compare diseases networks [9], identify physical designs of circuits[39], and discover molecules with similar properties[31, 3, 25]. In order to measure similarity between graphs effectively, several definitions of distance or similarity have been proposed[26, 14, 13, 16, 15]. For example, graph edit distances are the minimum cost for transforming one network to another by the distortion of nodes and edges[21]. These definitions only pay attention to the similarities of the nodes and edges but lacks the information of topological structures of the networks. For the purpose of addressing this limitation, frequency subgraph mining algorithms[41], graph kernels[36], and methods based on moments[32] have been proposed. Moreover, various distance between spectrums are used to measure similarity of graphs[23, 43, 28, 12, 30, 1, 44]. However, practically it is almost impossible to find the spectrum of a large-scale graph. Recently many different approaches are proposed for efficient algorithms [35, 2, 37]. However, these methods are not scalable to large-scale graphs containing millions of edges, which are common in today's applications. As a result, effective and scalable methods for large-scale graphs comparison are urgently needed.

[*]Corresponding author
*Email addresses:* hayoung79choi@gmail.com (Hayoung Choi), hosoolee@skku.edu (Hosoo Lee), shenyf@shanghaitech.edu.cn (Yifei Shen ), shiym@shanghaitech.edu.cn (Yuanming Shi )

[1]This article is partially based on preliminary results published in the proceeding of IEEE International Symposium on Information Theory from June 17 to 22, 2018 (ISIT2018Vail).

In this paper, we propose a novel similarity measure for comparing large-scale graphs. We consider the adjacency matrix of the graph as a real random variable on the algebraic probability space with the proposed state. We show that the spectral distribution of a Hermitian matrix in a given state can be expressed as a unique discrete probability measure. Then we propose an efficient and scalable method to measure the similarity between large-scale graphs based on the spectral distribution of the corresponding adjacency matrix in the given state. Specifically, we compute the corresponding positive semidefinite moment matrix whose entries consist of the first few number of moments of the spectrum distribution. Our proposed distance between graphs is obtained by a distance between the moment matrices. We show that this distance is graph invariant and substructure invariant. Moreover, it is scalable to extremely massive graphs and highly parallelable. Numerical simulations demonstrate that our proposed distance not only has better performance over the competing methods, but also outperforms the state-of-art method in collaboration network classification.

## 2. Background and Preliminary

Denote $\mathbb{N}$ (resp. $\mathbb{N}_+$) the set of nonnegative (resp. positive) integer numbers. Let $M_{m \times n} := M_{m \times n}(\mathbb{C})$ be a set of all $m \times n$ matrices with entries in the field $\mathbb{C}$ of complex numbers. We simply denote as $M_n := M_{n \times n}$. We equip on $M_{m \times n}$ with the inner product defined as

$$\langle A, B \rangle := \text{tr}(A^* B) = \sum_{i,j=1}^{m,n} \overline{a_{ij}} b_{ij},$$

for $A = [a_{ij}], B = [b_{ij}] \in M_{m \times n}$, where $A^* = \bar{A}^T$ is a complex conjugate transpose of $A$. The inner product naturally gives us an $\ell_2$ norm, known as the Frobenius norm and Hilbert-Schmidt norm,

defined by
$$\|A\|_2 = [\operatorname{tr}(A^*A)]^{1/2}.$$

The operator norm of $A \in M_n$ is defined as
$$\|A\| := \max_{\|x\|_2=1} \|Ax\|_2.$$

Note that
$$\|A\|_2 = \left[\sum_{i=1}^n \sigma_i^2(A)\right]^{1/2} \quad \text{and} \quad \|A\| = \sigma_1(A),$$

where $\sigma_1(A) \geq \cdots \geq \sigma_n(A)$ are (non-negative) singular values of $A$ in decreasing order.

2.1. Graph

Let $V$ be the set of vertices, and let $\{x,y\}$ denote the edge connecting two points $x,y \in V$. We say that two vertices $x,y \in V$ are *adjacent* if $\{x,y\} \in E$, denoted by $x \sim y$. A graph $G = (V,E)$ is called *finite* if $V$ is a finite set. Otherwise, it is called *infinite*. In general $E$ may contain *loops* which means that $x = y$. In this paper we consider a finite undirected graph with no loops. The *degree* of a vertex $x \in V$ is defined by $\deg(x) = |\{y \in V : y \sim x\}|$. Two graphs $G = (V,E)$ and $G' = (V',E')$ are *isomorphic* if there is a bijection $f : V \longrightarrow V'$ such that $u \sim v \iff f(u) \sim f(v)$, denoted by $G \cong G'$. For $m \in \mathbb{N}$, a finite sequence of vertices $x_0, x_1, \ldots, x_m \in V$ is called a *walk of length $m$* if $x_0 \sim x_1 \sim \cdots \sim x_m$, where some of $x_0, x_1, \ldots, x_m$ may coincide. A graph $G = (V,E)$ is *connected* if every pair of distinct vertices $x,y \in V$ ($x \neq y$) is connected by a walk. If there is a walk connecting two distinct vertices $x,y \in V$, the *graph distance* between $x$ and $y$ is the minimum length of a walk connecting $x$ and $y$, denoted by $\partial(x,y)$. For graphs $G_i = (V_i, E_i)$, $i = 1,2$ with $V_1 \cap V_2 = \emptyset$, the *direct sum* of $G_1$ and $G_2$ is defined as $G = (V_1 \cup V_2, E_1 \cup E_2)$, denoted by $G = G_1 \sqcup G_2$. Without loss of generality we assume that $V = \{1, 2, \ldots, n\}$. The *adjacency matrix* of a graph $G = (V,E)$ is a $n \times n$ matrix $A \in \{0,1\}^{n \times n}$ where $A_{ij} = 1$ if and only if $\{i,j\} \in E$ for all $i,j \in V$. Any graph $G$ can be represented by an adjacency matrix. Every permutation $\pi : \{1,2,\ldots,n\} \longrightarrow \{1,2,\ldots,n\}$ is associated with a corresponding *permutation matrix* $P$. The matrix operator $P$ left multiplied to matrix $A$ rearranges the rows according to $\pi$ which right multiplication with $P$ rearranges columns of the matrix $A$. Given an adjacency matrix $A$, graphs corresponding to adjacency matrix $A$ and $PAP^\top$ are isomorphic for any permutation matrix $P$, i.e., they represent the same graph structure. A property of graph is called *graph invariant* if the property does not change under the transformation of reordering of vertices. Note that the adjacency matrix of a graph includes the full information about a graph. For $x, y \in V$ and $m \in \mathbb{N}$ let $W_m(x,y)$ denote the number of walks of length $m$ connecting $x$ and $y$. Remark that $W_0(x,y) = 0$ if $x \neq y$ and $W_0(x,y) = 1$ if $x = y$.

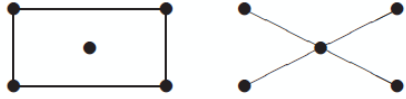

Figure 1: Cospectral graphs: $C_4 \cup K_1$ and $S_5$

**Theorem 2.1.** *Let $G = (V, E)$ be a graph and $A$ the adjacency matrix. Then we have*

$$(A^m)_{ij} = W_m(i, j)$$

*for all $i, j \in V$ and $m \in \mathbb{N}$.*

Let $\mathscr{A}(G)$ be the unital algebra generated by $A$ (the algebra generated by $A$ and the identity matrix $I = A^0$ ), i.e.,

$$\mathscr{A}(G) = \{f(A) : f \in \mathbb{C}[x]\},$$

where $\mathbb{C}[x]$ is the set of all polynomials with complex coefficients. Moreover, the involution is defined by $(cA^m)^* = \bar{c}A^m$ for $c \in \mathbb{C}$. Then $\mathscr{A}(G)$ becomes a unital $*$-algebra. We call $\mathscr{A}(G)$ *adjacency algebra* of $G$.

**Lemma 2.2.** *Let $G = (V, E)$ be a graph and $A$ the adjacency matrix. If there is a pair of vertices $x, y \in V$ such that $\partial(x, y) = d \geq 0$, then $I, A, \ldots, A^d$ are linearly independent and $\dim \mathscr{A}(G) \geq d+1$.*

For a finite connected graph $G = (V, E)$ the *diameter* is defined by

$$\mathrm{diam}(G) = \max\{\partial(x, y) : x, y \in V\}.$$

**Proposition 2.3.** *For a connected graph $G = (V, E)$ we have*

$$\dim \mathscr{A}(G) \geq \mathrm{diam}(G) + 1.$$

**Proposition 2.4.** *For a finite graph $G$ let $s(G)$ denote the number of distinct eigenvalue of $G$. Then we have $s(G) = \dim \mathscr{A}(G)$.*

**Corollary 2.5.** *For a connected finite graph $G$ we have*

$$s(G) = dim\mathscr{A}(G) \geq \mathrm{diam}(G) + 1.$$

It is clear that if $G \cong G'$ then the corresponding eigenvalues of the adjacency matrices are identical. However, in general the converse is not true.

*Cospectral graphs*, also called *isospectral graphs*, are graphs that share the same graph spectrum. The smallest pair of cospectral graphs is the graph union $C_4 \cup K_1$ and star graph $S_5$, illustrated in Figure 1. This is known that it is a unique pair of cospectral graphs among 34 non-isomorphic graphs on 5 vertices. Both have a common characteristic polynomial, $x^3(x-2)(x+2)$. For more examples for small graphs, see [42] and for more information about cospectral graphs see [7, 24, 20]. There are many attempts to distinguish cospectral graphs by means of another matrices.

2.2. *Quantum Probability*

For proof of each theorem and proposition, see [33] and references therein. To measure distance between two graphs we propose to compare the spectra of their adjacency matrices. Since the computational cost of finding all eigenvalues for large-scale graphs is very expensive, an alternative approach is brought by the idea of spectral distribution based on quantum probability theory.

**Definition 2.6.** *Let $\mathscr{A}$ be a unital $*$-algebra over the complex number field $\mathbb{C}$ with the multiplication unit $1_{\mathscr{A}}$. A function $\varphi : \mathscr{A} \longrightarrow \mathbb{C}$ is called a* state *on $\mathscr{A}$ if*

$$(i) \ \varphi \ \text{is linear}; \quad (ii) \ \varphi(a^*a) \geq 0; \quad (iii) \ \varphi(1_{\mathscr{A}}) = 1.$$

*The pair $(\mathscr{A}, \varphi)$ is called an algebraic probability space.*

**Proposition 2.7.** *A state $\varphi$ on a unital $*$-algebra $\mathscr{A}$ is a $*$-map, i.e., $\varphi(a^*) = \overline{\varphi(a)}$.*

**Definition 2.8.** *Let $(\mathscr{A}, \varphi)$ be an algebraic probability space. An element $a \in \mathscr{A}$ is called an* algebraic random variable *or a* random variable *for short. A random variable $a \in \mathscr{A}$ is called* real *if $a = a^*$.*

For a random variable $a \in \mathscr{A}$ the quantity of the form:

$$\varphi(a^{\varepsilon_1} a^{\varepsilon_2} \cdots a^{\varepsilon_m}), \quad \varepsilon_1, \varepsilon_2, \ldots, \varepsilon_m \in \{1, *\},$$

is called a *mixed moment* of order $m$. Statistical properties of an algebraic random variable are determined by its mixed moments. For a real random variable $a$ in $\mathscr{A}$ the mixed moments are reduced to the *moment sequence*:

$$\varphi(a^m), \quad m = 0, 1, 2, \ldots,$$

where $\varphi(a^m)$ is called the *m-th moment* of $a$. By definition $\varphi(a^0) = 1$.

For a real random variable $a = a^*$, a *moment matrix* with degree $n$ is defined as

$$\mathcal{M}_n := \begin{bmatrix} \varphi(a^0) & \varphi(a^1) & \cdots & \varphi(a^n) \\ \varphi(a^1) & \varphi(a^2) & \cdots & \varphi(a^{n+1}) \\ \vdots & \vdots & \ddots & \vdots \\ \varphi(a^n) & \varphi(a^{n+1}) & \cdots & \varphi(a^{2n}) \end{bmatrix}. \tag{2.1}$$

**Definition 2.9.** *Two algebraic random variables $a$ in $(\mathscr{A}, \varphi)$ and $b$ in $(\mathscr{B}, \psi)$ are* moment equivalent, *denoted by $a \stackrel{\circ}{=} b$, if their mixed moments coincide, i.e., if*

$$\varphi(a^{\varepsilon_1} a^{\varepsilon_2} \cdots a^{\varepsilon_m}) = \psi(b^{\varepsilon_1} b^{\varepsilon_2} \cdots b^{\varepsilon_m})$$

*for all $\varepsilon_1, \varepsilon_2, \ldots, \varepsilon_m \in \{1, *\}$ and $m \in \mathbb{N}$.*

Remark that for real random variables $a$ and $b$ it holds that $a \stackrel{\circ}{=} b$ if and only if $\varphi(a^m) = \psi(b^m)$ for all $m \in \mathbb{N}_0$.

Let $\mathfrak{B}(\mathbb{R})$ denote the set of all probability measures having finite moments of all orders.

**Theorem 2.10.** *Let $(\mathscr{A}, \varphi)$ be an algebraic probability space. For a real random variable $a = a^* \in \mathscr{A}$ there exists a probability measure $\mu \in \mathfrak{B}(\mathbb{R})$ such that*

$$\varphi(a^k) = \int_{\mathbb{R}} x^k \mathrm{d}\mu(x) \quad \textit{for all } k \in \mathbb{N}_0. \tag{2.2}$$

*Such $\mu$ is called the* spectral distribution *of $a$ in $\varphi$.*

It is noted that $M_n$ with the usual operators is a unital $*$-algebra. Recall that a matrix $\rho \in M_n$ is called a *density matrix* if it is positive semidefinite and $\operatorname{tr} \rho = 1$.

**Definition 2.11.** *For $A = [a_{ij}] \in M_n$, the following are states on $M_n$, implying that $(M_n, \varphi)$ is an algebraic probability space.*

(1) **(Normalized trace)** *The* normalized trace *is defined by*

$$\varphi_{\mathrm{tr}}(A) = \frac{1}{n} \operatorname{tr}(A) = \frac{1}{n} \sum_{i=1}^{n} a_{ii}.$$

(2) **(Vector state)** *For a unit vector $\xi \in \mathbb{C}^n$, we define*

$$\varphi_\xi(A) = \langle \xi, A\xi \rangle, \quad A \in M_n,$$

*where $\langle \cdot, \cdot \rangle$ is the usual inner product in $\mathbb{C}^n$. Such a state is called a* vector sate *with the state vector $\xi$.*

(3) **(Density matrix state)** *For each density matrix $\rho \in M_n$ we define*

$$\varphi_\rho(A) = \operatorname{tr}(\rho A), \quad A \in M_n.$$

*Such a state is called a* density matrix sate *with the density matrix $\rho$.*

**Proposition 2.12.** *For any state $\varphi$ on $M_n$ there exists a unique density matrix $\rho$ such that $\varphi = \varphi_\rho$.*

## 3. Main results

This following is the well-known result.

**Lemma 3.1.** *Let $A \in M_n$ have distinct eigenvalues $\lambda_1, \lambda_2, \ldots, \lambda_s$ and let*

$$q(t) = (t - \lambda_1)(t - \lambda_2) \cdots (t - \lambda_s).$$

*Then $A$ is diagonalizable if and only if $q(A) = 0$.*

**Theorem 3.2.** *Let $\{s_k\}$ be a real sequence and let*

$$\mathcal{H}_n = \begin{bmatrix} s_0 & s_1 & \cdots & s_n \\ s_1 & s_2 & \cdots & s_{n+1} \\ \vdots & \vdots & \ddots & \vdots \\ s_n & s_{n+1} & \cdots & s_{2n} \end{bmatrix}. \tag{3.1}$$

*be the Hankel matrix. If $\det(\mathcal{H}_n) > 0$ for all $n < s$ and $\det(\mathcal{H}_n) = 0$ for all $n \geq s$. Then there exists unique discrete measure, $\mu$, with $s$ number of point mass such that there exists unique discrete measure, $\mu$, with $s$ number of point mass such that*

$$s_k = \int_{\mathbb{R}} x^k d\mu \quad \text{for all } k \in \mathbb{N}.$$

*Proof.* See [4, Theorem 1.1]. □

A Hermitian matrix $A \in M_n$ can be regarded as a real random variable in the algebraic probability space $(M_n, \varphi_\xi)$ with a vector state $\varphi_\xi$, by Theorem 2.10 it follows that there exists the spectral distribution of $A$ in $\varphi_\xi$ such that

$$\varphi_\xi(A^k) = \xi^* A^k \xi = \int_{\mathbb{R}} x^k d\mu(x) \quad \text{for all } k \in \mathbb{N}. \tag{3.2}$$

In the following theorem we provide an explicit form of such measure.

Denote the Dirac measure at $\lambda$ as $\delta_\lambda$. (i.e., $\delta_\lambda(S) = 1$ if $\lambda \in S$ and $\delta_\lambda(S) = 0$ if $\lambda \notin S$). The support of measure, $\mu$, is denoted by $supp(\mu)$. A measure $\mu$ is called a *measure with $n_0$ mass point* if $|supp(\mu)| = n_0$.

**Theorem 3.3.** *Let $(M_n, \varphi_\xi)$ be the algebraic probability space with a vector state $\varphi_\xi$ and let $A \in M_n$ be a Hermitian matrix whose all eigenvalues are distinct, $\lambda_1, \ldots, \lambda_n$. There exists a unique probability discrete measure $\mu$ such that*

$$\varphi_\xi(A^k) = \int_{\mathbb{R}} x^k d\mu(x) \quad \text{for all } k \in \mathbb{N}. \tag{3.3}$$

*Furthermore, the measure has an explicit form $\mu = \sum_{i=1}^n \omega_i \delta_{\lambda_i}$. Conversely, for a probability discrete measure $\mu = \sum_{i=1}^n \omega_i \delta_{\lambda_i}$, there exists a Hermitian matrix $A \in M_n$ with eigenvalues $\lambda_1, \ldots, \lambda_n$ such that $A$ holds the equality (3.3).*

*Proof.* ($\Longrightarrow$) Let $A \in \mathbb{C}^{n \times n}$ be a Hermitian matrix. By Spectral Theorem, $A$ can be diagonalized by a unitary matrix $U$. That is, $A = UDU^*$. Put $v = U^*\xi = [v_1, \ldots, v_n]^\top \in \mathbb{C}^n$, $\omega_i = |v_i|^2$ for all $i = 1, \ldots, n$, and $\mu = \sum_i^n \omega_i \delta_{\lambda_i}$. Then the $k$th moment of $A$ is

$$\varphi_\xi(A^k) = \xi^* A^k \xi = v^* D^k v = \sum_{i=1}^n \omega_i \lambda_i^k = \int_\mathbb{R} x^k \mathrm{d}\mu.$$

Since it holds that

$$\int_\mathbb{R} \mathrm{d}\mu = \sum_{i=1}^n \omega_i = \sum_{i=1}^n v^* v = \sum_{i=1}^n \xi^* U U^* \xi = 1,$$

the measure $\mu$ is a probability measure.

($\Longleftarrow$) Let $\mu = \sum_{i=1}^n \omega_i \delta_{\lambda_i}$ with $\omega_i \geq 0$ and $\lambda_i \in \mathbb{R}$ for all $i \in \mathbb{N}$, and $\sum_{i=1}^n \omega_i = 1$. Let $D$ be the $n \times n$ diagonal matrix whose diagonal entries are $\lambda_1, \lambda_2, \ldots, \lambda_n$. Let $v = [\sqrt{\omega_1} \ldots \sqrt{\omega_n}]^\top$. Since $v$ and $\xi$ both are unit vectors, there exists a unitary matrix $U$ such that $Uv = \xi$. Take $A = UDU^*$. Then $A$ holds the equality (3.3). $\square$

Note that $\langle u_i, \xi \rangle = \cos \theta_i$ such that $\theta_i$ is the angle between $u_i$ and $\xi$, where $u_i$ is $i$th column vector of $U$. Since $\sum \omega_i = 1$, it holds that $\sum |\cos \theta_i|^2 = 1$. So, $\cos \theta_i$ are the direction cosines of $\xi$ with respect to orthonormal eigenvectors, $u_1, u_2, \ldots, u_n$. Especially, if $\xi = u_i$, then the spectral distribution is $\mu = \delta_{\lambda_i}$.

**Corollary 3.4.** *Let $(M_n, \varphi_\mathrm{tr})$ be the algebraic probability space with the normalized trace state $\varphi_\mathrm{tr}$ and let $A \in M_n$ be a Hermitian matrix whose all eigenvalues, $\lambda_1, \ldots, \lambda_n$, are distinct. An explicit form of the unique probability discrete measure $\mu$ such that $\varphi_\mathrm{tr}(A^k) = \int_\mathbb{R} x^k \mathrm{d}\mu(x)$ for all $k \in \mathbb{N}$ is $\mu = \frac{1}{n} \sum_{i=1}^n \delta_{\lambda_i}$.*

Remark that while the spectral distribution of a Hermitian matrix $A$ in the normalized trace state includes only information about eigenvalues for $A$, the spectral distribution of $A$ in the vector state includes information about the corresponding eigenvectors as well as eigenvalues for $A$.

Now we generalize for any Hermitian matrices.

**Theorem 3.5.** *Let $(M_n, \varphi_\xi)$ be the algebraic probability space with a vector state $\varphi_\xi$ and let $A \in M_n$ be a real random variable. Let $\lambda_1, \lambda_2, \ldots, \lambda_s$ with respective multiplicities $n_1, \ldots, n_s$, and let $\Lambda = \lambda_1 I_{n_1} \oplus \cdots \oplus \lambda_s I_{n_s}$. Supposed that $U = [U_1 \ U_2 \ \cdots \ U_s] \in M_n$ is unitary matrix such that $A = U \Lambda U^*$ and for each $i = 1, 2, \ldots, s$,*

$$U_i = [u_1^{(i)} \ u_2^{(i)} \ \cdots \ u_{n_i}^{(i)}] \in M_{n \times n_i},$$

where $u_1^{(i)}, \ldots, u_{n_i}^{(i)}$ are the corresponding unit eigenvectors of $\lambda_i$. Then there uniquely exists a probability discrete measure, $\mu = \sum_{i=1}^{s} \omega_i \delta_{\lambda_i}$, such that

$$\varphi_\xi(A^k) = \int_\mathbb{R} x^k \mathrm{d}\mu(x) \quad \text{for all } k \in \mathbb{N}. \tag{3.4}$$

Furthermore, $\omega_i = \sum_{j=1}^{n_i} |\cos\theta_j^{(i)}|^2$ where $\theta_j^{(i)}$ is the angle between $u_j^{(i)}$ and $\xi$.

*Proof.* Note that since $A$ is a Hermitian matrix, by Spectral Theorem, $A$ can be diagonalized by an unitary matrix. Let $\omega_i = \sum_{j=1}^{n_i} |\langle u_j^{(i)}, \xi \rangle|^2$ and $\mu = \sum_{i=1}^{s} \omega_i \delta_{\lambda_i}$. Then it follows that for each $k \in \mathbb{N}$

$$\varphi_\xi(A^k) = \xi^* U \Lambda^k U^* \xi = \xi^* U_1 \lambda_1^k I_{n_1} U_1^* \xi \oplus \cdots \oplus \xi^* U_s \lambda_s^k I_{n_s} U_s^* \xi = \sum_{i=1}^{s} \omega_i \lambda_i^k = \int_\mathbb{R} x^k \mathrm{d}\mu(x).$$

Since it holds that

$$\int_\mathbb{R} d\mu = \sum_{i=1}^{s} \omega_i = \sum_{i=1}^{s} \sum_{j=1}^{n_i} |\langle u_j^{(i)}, \xi \rangle|^2 = \sum_{i=1}^{s} |U_i^* \xi|^2 = \xi^* U U^* \xi = 1$$

the measure $\mu$ is a probability measure. Since $u_j^{(i)}$ for all $i, j$ is a unit vector and $\xi$ is also a unit vector, it follows that

$$\omega_i = \sum_{j=1}^{n_i} |\langle u_j^{(i)}, \xi \rangle|^2 = \sum_{j=1}^{n_i} |\cos\theta_j^{(i)}|^2,$$

where $\theta_j^{(i)}$ is the angle between $u_j^{(i)}$ and $\xi$.

(**Uniqueness**) (To apply Theorem 3.2, we have to show that $\det(\mathcal{M}_j)$ is nonnegative.) For the moment sequence $\{\varphi_\xi(A^k)\}_k$, the corresponding moment matrix $\mathcal{M}_j$ with degree $j$ is defined as

$$\mathcal{M}_j := \begin{bmatrix} \varphi_\xi(A^0) & \varphi_\xi(A^1) & \cdots & \varphi_\xi(A^j) \\ \varphi_\xi(A^1) & \varphi_\xi(A^2) & \cdots & \varphi_\xi(A^{j+1}) \\ \vdots & \vdots & \ddots & \vdots \\ \varphi_\xi(A^j) & \varphi_\xi(A^{j+1}) & \cdots & \varphi_\xi(A^{2j}) \end{bmatrix} = \begin{bmatrix} \xi^* I \xi & \xi^* A \xi & \cdots & \xi^* A^j \xi \\ \xi^* A \xi & \xi^* A^2 \xi & \cdots & \xi^* A^{j+1} \xi \\ \vdots & \vdots & \ddots & \vdots \\ \xi^* A^j \xi & \xi^* A^{j+1} \xi & \cdots & \xi^* A^{2j} \xi \end{bmatrix}.$$

Since for any $n$

$$\mathcal{M}_n = \begin{bmatrix} I\xi & A\xi & \ldots & A^j\xi \end{bmatrix}^* \begin{bmatrix} I\xi & A\xi & \ldots & A^j\xi \end{bmatrix},$$

$\mathcal{M}_n$ is positive semidefinite and $\det(\mathcal{M}_j) \geq 0$.

Let $\lambda_1, \lambda_2, \ldots, \lambda_s$ be the distinct eigenvalues of $A$ and

$$q(t) = (t - \lambda_1)(t - \lambda_2) \cdots (t - \lambda_s) = \sum_{j=1}^{s} c_j t^j$$

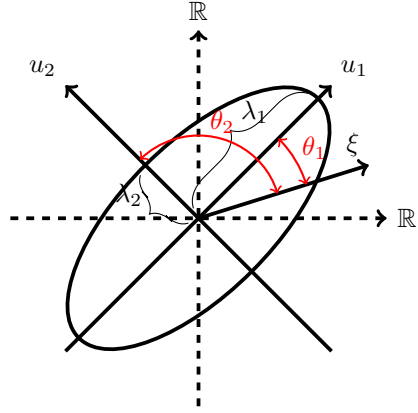

Figure 2: For a symmetric matrix $A \in M_{2\times 2}(\mathbb{R})$ and a unit vector $\xi \in \mathbb{R}^2$, there exists the probability discrete measure $\mu = \cos^2\theta_1 \delta_{\lambda_1} + \cos^2\theta_2 \delta_{\lambda_2}$ such that $\xi^\top A\xi = \int_\mathbb{R} x^k d\mu$ for all $k \in \mathbb{N}$.

Since $A$ is a Hermitian matrix, by Lemma 3.1 it follows that $q(A) = \sum_{j=1}^{s} c_j A^j = 0$. Then it follows that

$$0 = \langle \xi, q(A)\xi \rangle = \xi^* \sum_{j=1}^{s} c_j A^j \xi = \sum_{j=1}^{s} c_j \varphi_\xi(A^j).$$

So, it is clear that $\det(\mathcal{M}_j) = 0$ for $j \geq s$ and $\det(\mathcal{M}_j) > 0$ for $j < s$. By Theorem 3.2, there exists a unique discrete measure with $s$ number of point mass. Therefore, $\mu$ is a unique probability discrete measure. □

**Example 3.6.** *Consider $2 \times 2$ symmetric matrix $A \in M_2$ as*

$$A = \begin{bmatrix} 2 & 1 \\ 1 & 2 \end{bmatrix} = \begin{bmatrix} 1/\sqrt{2} & -1/\sqrt{2} \\ 1/\sqrt{2} & 1/\sqrt{2} \end{bmatrix} \begin{bmatrix} 3 & 0 \\ 0 & 1 \end{bmatrix} \begin{bmatrix} 1/\sqrt{2} & 1/\sqrt{2} \\ -1/\sqrt{2} & 1/\sqrt{2} \end{bmatrix}.$$

*Let $\xi = [\xi_1 \ \xi_2]^\top$ be a unit vector in $\mathbb{R}^2$. Then by Theorem 3.3, there exists a unique probability measure $\mu = \omega_1 \delta_{\lambda_1} + \omega_2 \delta_{\lambda_2}$ such that*

$$\varphi_\xi(A^k) = \xi^\top A^k \xi = \int_\mathbb{R} x^k d\mu(x) \quad \text{for all } k \in \mathbb{N}.$$

*It is easy to check a closed form of the measure is $\mu = (1/2 + \xi_1\xi_2)\delta_3 + (1/2 - \xi_1\xi_2)\delta_1$. The weights $\omega_1, \omega_2$ depends on $\xi$. Especially, when $\xi = [1/\sqrt{2} \ \ 1/\sqrt{2}]^\top$, the spectral distribution of $A$ is $\mu = \delta_3$. When $\xi = [-1/\sqrt{2} \ \ 1/\sqrt{2}]^\top$, the spectral distribution of $A$ is $\mu = \delta_1$ (see Figure 2).*

Remark that the spectral distribution includes information not only about the eigenvalues of $A$, but also about the corresponding eigenvectors. Since $\omega_i$ depends on $\xi$ for all $i$, the spectral distribution of a given Hermitian matrix $A \in M_n$ depends on a unit vector $\xi \in \mathbb{C}^n$. However, since the eigenvalues do not change, the Dirac measures $\delta_{\lambda_i}$ do not change. The only weights $\omega_i$ depends on a unit vector $\xi \in \mathbb{C}^n$.

**Lemma 3.7.** *Let $A$ be a $n \times n$ Hermitian matrix and $\xi$ be a unit column vector in $\mathbb{C}^n$. Then $\varphi_\xi(A^k) = \lambda^k$ for all $k \in \mathbb{N}$ if and only if $A\xi = \lambda\xi$.*

*Proof.* ($\Longleftarrow$) Since $A\xi = \lambda\xi$ implies $A^k\xi = \lambda^k\xi$ for all $k \in \mathbb{N}$, it follows that $\xi^* A^k \xi = \xi^* \lambda^k \xi = \lambda^k$.

($\Longrightarrow$) Since $\varphi_\xi(A^k) = \lambda^k$, by definition $\xi^* A^k \xi = \lambda^k$. Then by Theorem 3.5 it follows that $\sum_{i=1}^s \omega_i \lambda_i^k = \lambda^k$ for all $k \in \mathbb{N}$, where $\lambda_1, \ldots, \lambda_s$ are the distinct eigenvalues of $A$. By the uniqueness of the spectral distribution, $\lambda = \lambda_\ell$ for some $1 \leq \ell \leq s$. The following linear system

$$\begin{bmatrix} 1 & 1 & \cdots & \cdots & 1 \\ \lambda_1 & \lambda_2 & \cdots & \cdots & \lambda_s \\ \lambda_1^2 & \lambda_2^2 & \cdots & \cdots & \lambda_s^2 \\ \vdots & \vdots & \ddots & \ddots & \vdots \\ \lambda_1^{s-1} & \lambda_2^{s-1} & \cdots & \cdots & \lambda_s^{s-1} \end{bmatrix} \begin{bmatrix} \omega_1 \\ \omega_2 \\ \omega_3 \\ \vdots \\ \omega_s \end{bmatrix} = \begin{bmatrix} 1 \\ \lambda_\ell^1 \\ \lambda_\ell^2 \\ \vdots \\ \lambda_\ell^s \end{bmatrix}$$

has a unique solution $\omega_i = 0$ for all $i \neq \ell$ and $\omega_\ell = 1$. Thus, $\xi$ is the corresponding eigenvector of $\lambda_\ell$. $\square$

**Lemma 3.8.** *If Hermitian matrices $A, \tilde{A} \in M_n$ hold that $\varphi_{tr}(A^k) = \varphi_{tr}(\tilde{A}^k)$ for all $k = 1, 2, \ldots, n$, then $A$ and $\tilde{A}$ have same spectrums.*

*Proof.* Recall that a monic polynomial is a univariate polynomial in which the leading coefficient (the nonzero coefficient of highest degree) is equal to 1. Let $\lambda_i, \tilde{\lambda}_i$ be eigenvalues of $A, \tilde{A}$, respectively. Let $f$ and $\hat{f}$ be degree $m$ monic polynomial functions whose roots consist of eigenvalues of $A$ and $\tilde{A}$, respectively. Since $\sum_i (\lambda_i)^k = \sum_i (\tilde{\lambda}_i)^k$, $k = 1, 2, \ldots, n$, by Newton's identities it follows that the coefficients of two polynomials $f$ and $\hat{f}$ are identical. Thus, the roots of $f$ and $\hat{f}$ are identical. Therefore, $\lambda_i = \tilde{\lambda}_i$ for all $i$. $\square$

**Theorem 3.9.** *The following statement are equivalent.*

*(i) There exists a unique discrete measure $\mu$ with $n_0$ mass points such that $m_k = \int x^k d\mu$ for all $k \in \mathbb{N}$;*

*(ii) There is a Hermitian matrix $A \in \mathbb{C}^{m \times m}$ with $n_0$ distinct eigenvalues and a unit vector $\xi \in \mathbb{C}^m$ such that $m_k = \xi^* A^k \xi$ for all $k \in \mathbb{N}$;*

*(iii) $\mathcal{M}_n \geq 0$ for all $n \in \mathbb{N}$, and $\mathcal{M}_n > 0$ if and only if $n < n_0$.*

*Proof.* (i) $\Rightarrow$ (ii) Let $\mu = \sum_{i=1}^{n_0} \omega_i \delta_{\lambda_i}$ with $\omega_i \geq 0$ and $\lambda_i \in \mathbb{R}$ for all $i \in \mathbb{N}$, and $\sum_{i=1}^{n_0} \omega_i = 1$ and $\xi$ be a unit vector in $\mathbb{C}^n$. Let $D$ be the $n \times n$ diagonal matrix whose diagonal entries are $\lambda_1, \lambda_2, \ldots, \lambda_{n_0}, 0, \ldots, 0$. Let $v = [\sqrt{\omega_1} \cdots \sqrt{\omega_{n_0}} \, 0 \cdots 0]^\top$. Since $v$ and $\xi$ both are unit vectors, there exists a unitary matrix $U$ such that $Uv = \xi$. Let $A = UDU^*$. Then it follows that $m_k = \xi^* A^k \xi$ for all $k \in \mathbb{N}$.

(ii) $\Rightarrow$ (iii) Let $c = [c_0, c_1, \ldots, c_n]^*$ be a vector in $\mathbb{C}^{n+1}$. Then

$$c^* \mathcal{M}_n c = \sum_{i,j=0}^{n} m_{i+j} c_i^* c_j = \sum_{i,j=0}^{n} (e^* A^{i+j} e) c_i^* c_j = \left\| \sum_{i=0}^{n} c_i A^i e \right\|^2 \geq 0.$$

Since $c \in \mathbb{C}^{n+1}$ is arbitrary, $\mathcal{M}_n$ is positive semidefinite for all $n \in \mathbb{N}$. Since $A$ is Hermitian, the minimal polynomial is $q(x) := (x - \lambda_1)(x - \lambda_2) \cdots (x - \lambda_{n_0})$ were $\lambda_i$ are all eigenvalues. So, $\sum_{i=1}^{n_0} q_i A^i = 0$, implying $\|\sum_{i=0}^{n_0} q_i A^i e\| = 0$. Thus $\mathcal{M}_n$ is singular for $n \geq n_0$. Suppose that there exists a polynomial $r(x) := \sum_i^m r_i x^i$ with $m < n_0$ such that $r(A)e = 0$. Since $n_0$ is the degree of minimal polynomial, $r(x)$ is a zero function.

(iii) $\Rightarrow$ (i) See [4, Theorem 1.1]. $\square$

Denote the set of $n \times n$ permutation matrices as $\mathcal{S}$. Denote the identity matrix as $I$ and the matrix whose all entries are 1 as $J$.

**Definition 3.10.** *Let $(M_n, \varphi)$ be an algebraic probability space. A state $\varphi$ is called permutationally invariant on $M_n$ if*

$$\varphi(A) = \varphi(P^\top A P) \quad \text{for all } A \in M_n, \ P \in \mathcal{S}. \tag{3.5}$$

**Lemma 3.11.** *A necessary and sufficient condition that $\varphi$ is permutationally invariant is that there exists a density matrix $\rho$ such that $\varphi(A) = \text{tr}(\rho A)$ satisfying*

$$\rho = pI + qJ, \quad n(p+q) = 1, \quad p \geq 0, \quad p + qn \geq 0. \tag{3.6}$$

*Proof.* Recall that for any state $\varphi$ on $M_n$ there exists a unique density matrix $\rho$ such that $\varphi = \varphi_\rho$.

($\Longrightarrow$) If $\rho = pI + qJ$, then $\text{tr}(\rho P^\top A P) = \text{tr}(P^\top \rho P A) = \text{tr}(\rho A)$ for all $A \in M_n, \ P \in \mathcal{S}$.

($\Longleftarrow$) Let $\rho = [\rho_{ij}]$ be a density matrix. Since $\text{tr}(\rho P^\top A^\top P) = \text{tr}(\rho A^\top)$ for all

$$A = \begin{bmatrix} a_{11} & a_{12} & 0 & \cdots & 0 \\ a_{21} & a_{22} & 0 & \cdots & 0 \\ 0 & 0 & 0 & \cdots & 0 \\ \vdots & \vdots & \vdots & \ddots & \vdots \\ 0 & 0 & 0 & \cdots & 0 \end{bmatrix} \in M_n,$$

it holds $\rho_{11}a_{11} + \rho_{12}a_{12} + \rho_{21}a_{21} + \rho_{22}a_{22} = \rho_{ii}a_{11} + \rho_{ij}a_{12} + \rho_{ji}a_{21} + \rho_{jj}a_{22}$ for all $i, j$. Since $a_{11}, a_{12}, a_{21}, a_{22}$ are arbitrary, $\rho_{ii} = \rho_{jj}$ for all $i, j$ and $\rho_{ij} = \rho_{k\ell}$ for all $i \neq j$, $k \neq \ell$. So, $\rho$ is of the form $\rho = pI + qJ$. Note that the eigenvalues of $pI + qJ$ are $p$ and $p + qn$. Since $\text{tr}(pI + qJ) = 1$ and $pI + qJ \geq 0$, it follows that $n(p+q) = 1$, $p \geq 0$, and $p + qn \geq 0$.

$\square$

**Theorem 3.12.** *Let $A$ be the adjacency matrix of a given graph $G$. Then the $k$-th moment of $A$ in a permutationally invariant state is a graph invariant.*

Let $A$ be the adjacency matrix of a given graph $G$. By Theorem 2.1, it is easy to check that $\varphi_{\text{tr}}(A^k)$ is the average of closed walks of length $k$ in $G$. Denote the $n$ dimensional all-ones column vector by $1_n$, and denote $1_n/\|1_n\|$ by $e$. From now on, the vector state with the state vector $\xi = e$ will be mainly used to compare two graphs. Specifically, the state $\varphi_e : M_n \to \mathbb{C}$ is defined by

$$\varphi_e(A) = \langle e, Ae \rangle \tag{3.7}$$

for all $A \in M_n$. Then it is clear that $\varphi_e$ is a state on $M_n$, implying that $(M_n, \varphi_e)$ is an algebraic probability space. Note that it holds

$$\varphi_e(A^k) = \frac{1}{n} \langle 1_n, A^k 1_n \rangle = \mathbb{E}[A^k 1_n],$$

where $\mathbb{E}(v) = \dfrac{1}{n} \sum_{i=1}^{n} v_i$ is the average of entries of vector $v$. Since the value of $(A^k)_{i,j}$ is equal to the number of walks of length $k$ from vertex $i$ to vertex $j$ and $A^k 1_n$ is the column vector whose $i$-th entry is equal to the sum of the number of all walks of length $k$ from the vertex $i$, $\varphi_e(A^k)$ is the average of the the sum of the number of all walks of length $k$ from each vertex.

**Proposition 3.13.** *If $\varphi_{\text{tr}}(A) = \varphi_{\text{tr}}(B)$ and $\varphi_e(A) = \varphi_e(B)$ for all $k \in \mathbb{N}$, then $\varphi(A) = \varphi(B)$ for all permutationally invariant state $\varphi$.*

*Proof.* Note that $\varphi_{\text{tr}}(A) = \frac{1}{n} \text{tr}(IA)$ and $\varphi_e(A) = \frac{1}{n} \text{tr}(JA)$ for all $A \in M_n$. $\square$

In other words, if the averages of closed walks and all walks of length $k$ in two graphs are idetical for all $k$, then their adjacency matrices are moment equivalent in $(M_n, \varphi)$ for any permutationally invariant state $\varphi$. Especially, if two Hermitian matrices $A, B \in M_n$ have distinct eigenvalues, respectively, and $\varphi_e(A^k) = \varphi_e(B^k)$ for all $k \in \mathbb{N}$, then for all permutation invariant state $\varphi$, we have $\varphi(A^k) = \varphi(B^k)$ for all $k \in \mathbb{N}$.

$\varphi_e$ has more properties

**Proposition 3.14.** *Let $A \in M_n$ be an adjacency matrix of a given graph. Then the following are true.*

(1) $\varphi_e(A) \leq \frac{1}{n} \sum_{i=1}^n \deg(i)^k$ for all $k \geq 1$,

(2) $\varphi_e(A^k) \leq \triangle^k$ for all $k \geq 1$ and $\varphi_e(A^k) \leq 2\varphi_e(A)\triangle^{k-1}$ for all $k \geq 2$,

(3) $(\varphi_e(A))^k \leq \varphi_e(A^k)$ for all $k \in \mathbb{N}$,

(4) $\varphi_e(A^{2a+b})\varphi_e(A^b) \leq \varphi_e(A^{2a+2b})$ for all $a, b \in \mathbb{N}$,

(5) $\varphi_e(A^{a+b})\varphi_e(A^{a+b}) \leq \varphi_e(A^{2a})\varphi_e(A^{2b})$ for all $a, b \in \mathbb{N}$,

where $\triangle$ is the maximum degree.

*Proof.* For (1) see [19]. For (2) see [10, Theorem 2]. For (3)–(5) see [40, Theorem 1-3]. □

Since $\varphi_e$ is a permutationally invariant state, the $k$-th moment of $A$ in $\varphi_e$ is a graph invariant. Then it holds the following.

**Theorem 3.15.** *Let $A$ be the adjacency matrix of a given graph $G$. Then the moment matrix $\mathcal{M}_n$ is a graph invariant.*

Hence, we will henceforth denote $\mathcal{M}_n$ as $\mathcal{M}_n(G)$ if a graph $G$ is given. $\mathcal{M}_n(G)$ is an informative representation for the given graph $G$. Indeed, $\mathcal{M}_n(G)$ includes information about the spectral properties of the adjacent matrix of $G$. $\mathbb{E}[Ae]$ is equal to $2|E|/|V|$. $\mathbb{E}[A^2 e]$ is equal to $2|E| + 2P_2$ where $P_2$ is the total number of distinct simple paths of length 2 (see [38, Lemma 1]). Remark that the variance $\mathbb{V}[Ae] = \mathbb{E}[A^2 e] - (\mathbb{E}[Ae])^2$.

Remark that the spectral distribution includes information not only about the eigenvalues of the adjacency matrix $A$, but also about the corresponding eigenvectors. To measure similarity between two large-scale graphs, we compare the spectral distributions of their adjacency matrices. There are various distances and divergences between two distributions such as Kullback-Leibler divergence, Bhattacharyya distance, etc (see [11, 29]). However, since large-scale graphs in real world have rich spectrums, to reconstruct the spectral distributions is impossible in practice. Instead, we can use moments of the distributions. In general, all the moments up to infinity are required to obtain a perfect reconstruction. However, the first few moments are only sufficient if the class of functions in which the reconstruction is sought is restricted appropriately. It has been mentioned in the literature that the most of the information about the measure is contained in the first few moments, and the higher-order ones providing only little additional information [18, 27, 22]. Since the moment matrix has sufficient information about the distribution, a distance between moment matrices can be calculated to measure a distance between two spectral distributions.

For two graphs $G$ and $\widetilde{G}$, we propose new distance between $G$ and $\widetilde{G}$ as a distance between the corresponding moment matrices, i.e.,

$$d(G, \widetilde{G}) := \delta(\mathcal{M}_m(G), \mathcal{M}_m(\widetilde{G})),$$

where $m \in \mathbb{N}$ is fixed and $\delta(\cdot, \cdot)$ is a distance between positive definite matrices.

**Theorem 3.16.** *For graphs $G, \tilde{G}, \hat{G}$,*

(a) *(Nonnegativity)* $d(G, \tilde{G}) \geq 0$,

(b) *(Identification)* $d(G, \tilde{G}) = 0$ *if* $G = \tilde{G}$,

(c) *(Symmetry)* $d(G, \tilde{G}) = d(\tilde{G}, G)$,

(d) *(Triangle Inequality)* $d(G, \hat{G}) \leq d(G, \tilde{G}) + d(\tilde{G}, \hat{G})$.

Recall that Theorem 3.9 states that the moment matrix in (2.1) is positive semidefinite matrix for all $n \in \mathbb{N}$. However, the corresponding moment matrix $\mathcal{M}_n$ can possibly be a singular positive semidefinite matrix, which is not a positive definite matrix. If $\mathcal{M}_m$ is positive definite for some $m$, then it is a point on the Riemannian manifold of positive definite matrices (see [4, Theorem 1.1]).

Denote the set of all $n \times n$ positive definite matrices as $\mathcal{P}^o$. There are various distances between two positive definite matrices such as Frobenius, Cholesky-Frobenius, J-divergence, Affine-invariant, Log-Frobenius [6]. The Frobenius norm $\|\cdot\|_2$ gives rise to the affine-invariant metric on $\mathcal{P}^o$ given by $\delta(A, B) = \|\log(A^{-1/2}BA^{-1/2})\|_2$ for any $A, B \in \mathcal{P}^o$. Then $\mathcal{P}^o$ is a Cartan-Hadamard manifold, a simply connected complete Riemannian manifold with non-positive sectional curvature. The geodesic curve has a parametrization $\gamma(t) = A^{1/2}(A^{-1/2}BA^{-1/2})^t A^{1/2}$, $0 \leq t \leq 1$, which is the unique geodesic from $A$ to $B$ (see [5]).

The computational results showed that the geodesic distance is a little bit better than other distances. In this paper, we use the geodesic distance. However, since the moment matrix for a graph with few vertices can be possibly singular positive semidefinite, we use the Frobenius distance instead. We remain which distance is the best in some sense for the future work.

**Definition 3.17.** *A property of graphs is called* sub-structure invariant *if the property of $G$ holds for $G \sqcup G \sqcup \cdots \sqcup G$.*

**Lemma 3.18.** *Let $G_i = (V_i, E_i)$ be a graph, $i = 1, 2$ with $V_1 \cap V_2 = \emptyset$. Then*

$$\mathcal{M}_n(G_1 \sqcup G_2) = \alpha \mathcal{M}_n(G_1) + (1-\alpha)\mathcal{M}_n(G_2),$$

*where $\alpha = |V_1|/(|V_1| + |V_2|)$.*

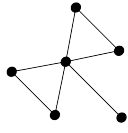
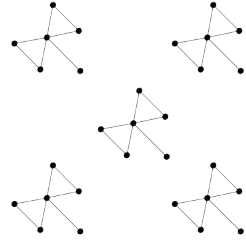

Figure 3: Graph $G$

Figure 4: Graph $\tilde{G} = G \sqcup G \sqcup G \sqcup G \sqcup G$

Two graphs have same sub-structure. Moreover, If $A$ and $\tilde{A}$ are considered as real algebraic random variables in $(\mathscr{A}(G), \varphi_e)$ and $(\mathscr{A}(\tilde{G}), \varphi_e)$, then two algebraic random variables $A$ and $\tilde{A}$ are moment equivalent.

Using the fact that $A_{Kn \times Kn} = I_{K \times K} \otimes A$ the preceding result can be extended easily as follows.

**Theorem 3.19.** Let $G_1, G_2, \ldots, G_K$ be given mutually disjoint graphs. Then

$$\mathcal{M}_n(G_1 \sqcup \ldots \sqcup G_K) = \sum_{j=1}^{K} \frac{\alpha_j}{\alpha} \mathcal{M}_n(G_j)$$

where $\alpha_j$ is the number of vertices in $G_j$ and $\alpha = \alpha_1 + \cdots + \alpha_K$.

Remark that if $m_1 = m_2 = \cdots = m_K$, then

$$\mathcal{M}_n(G_1 \sqcup \ldots \sqcup G_K) = \frac{1}{K}\left(\mathcal{M}_n(G_1) + \cdots + \mathcal{M}_n(G_K)\right).$$

Especially,

$$\mathcal{M}_n(G \sqcup \ldots \sqcup G) = \mathcal{M}_n(G) \quad \text{for all } n \in \mathbb{N}.$$

Specifically, $k$-th moments of adjacency matrices of $G$ and $G \sqcup G$ are same for each $k \in \mathbb{N}$, so their distributions are identical. If a graph consists of identical subgraphs, then the moment matrix of a given graph is equal to one of its subgraph. In other words, a moment matrix of graph can preserve information regardless of repetition of structure (see Figure 4). Thus, this property allows to use a subgraph to calculate distance between two large-scale networks.

**Corollary 3.20.** Let $A$ be the adjacency matrix of a given graph $G$. Then the $k$-th moment of $A$ in $\varphi$ is sub-structure invariant.

## 4. Cospectral graphs and various examples

The smallest pair of cospectral graphs is the graph union $C_4 \cup K_1$ and star graph $S_5$, illustrated in Figure 1. While the corresponding adjacency matrices are different, both have the same graph spectrum, $-2, 0, 0, 0, 2$. Let $A$ and $\tilde{A}$ be the adjacency matrix of $C_4 \cup K_1$ and $S_5$, respectively. If $A$ and $\tilde{A}$ are considered as real algebraic random variables in $(\mathscr{A}(C_4 \cup K_1), \varphi_{\mathrm{tr}})$ and $(\mathscr{A}(S_5), \varphi_{\mathrm{tr}})$, then two algebraic random variables $A$ and $\tilde{A}$ are moment equivalent, since

$$\varphi_{\mathrm{tr}}(A^k) = \frac{1}{n}\mathrm{tr}(A^k) = \frac{1}{n}\mathrm{tr}(\tilde{A}^k) = \varphi_{\mathrm{tr}}(\tilde{A}^k). \tag{4.1}$$

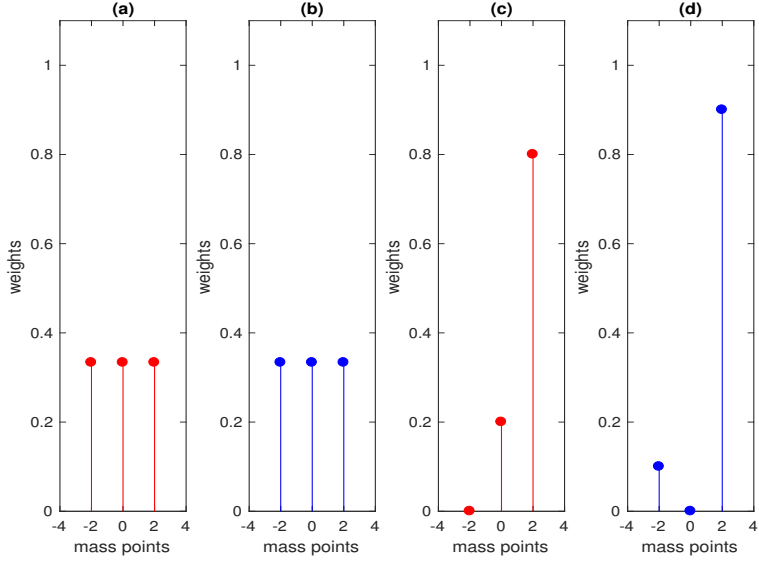

Figure 5: (a) and (b) are spectral distributions of $G_1$ and $G_2$ for trace state. (c) and (d) are spectral distributions of $G_1$ and $G_2$ for our proposed state. It shows that while (a) and (b) are identical, (c) and (d) are distinguishable.

That is, they have identical spectral distributions (see (a), (b) in Figure 5).

However, If $A$ and $\tilde{A}$ are considered as real algebraic random variables in $(\mathscr{A}(C_4 \cup K_1), \varphi_e)$ and $(\mathscr{A}(S_5), \varphi_e)$, then two algebraic random variables $A$ and $\tilde{A}$ are not moment equivalent. So, using the state $\varphi_e$ allows us to distinguish two graphs. Indeed, the moment matrices

$$\mathcal{M}_1(C_4 \cup K_1) = \begin{pmatrix} 1 & 1.6 \\ 1.6 & 3.2 \end{pmatrix}, \quad \mathcal{M}_1(S_5) = \begin{pmatrix} 1 & 1.6 \\ 1.6 & 4 \end{pmatrix}$$

are different, implying that each spectral distributions are different (see (c), (d) in Figure 5).

Fig. 6 is introduced in [35]. Three graphs have the same number of vertices and edges. Table 1 shows distances between the graphs based on Hamming distance, graph edit distance, D-measure, and our proposed distance. A good measure should return a higher distance value between $N1$ and $N3$ than between $N1$ and $N2$. Hamming distance and graph edit distance do not capture relevant topological differences. However, D-measure and our proposed measure perform a highly precise comparison.

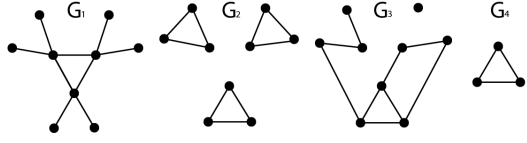

Figure 6: Three different networks with the same number of vertices and edges are shown in [35].

| Dissimilarity | H | GED | D-measure | Proposed measure |
|---|---|---|---|---|
| $d(G1, G2)$ | 12 | 6 | 0.252 | 14.0844 |
| $d(G1, G3)$ | 12 | 6 | 0.565 | 30.3974 |
| $d(G2, G3)$ | 12 | 6 | 0.473 | 16.3209 |

Table 1: (i) H: Hamming distance; (ii) GED: graph edit distance; (iii) D-measure: a recent proposed method in [35]; (iv) our propose method.

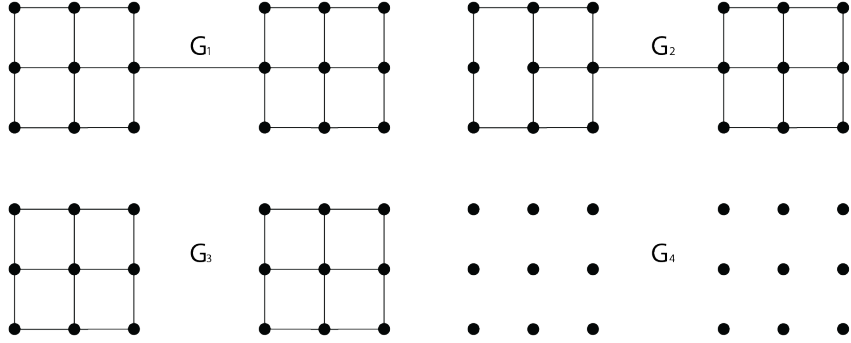

Figure 7: Five different networks with the same number of vertices in [43].

Five graphs $G_1$, $G_2$, $G_3$, $G_4$ and $G_5 = K_{18}$ on 18 vertices in Fig. 7 are considered in [43] to measure dissimilarity between them. The difference between $G_1$ and $G_2$ is only one edge. There is also only one edge difference between $G_1$ and $G_3$ but $G_3$ is not connected. While considering graphs as matrices or vectors, the distance between $G_1$ and $G_2$ is the same as between $G_1$ and $G_3$. However, $G_3$ is totally different as it is not connected. In [43] the following distance to compare graphs is proposed

$$d(G, G') = \sum_{i,j} \frac{(\lambda_i - \mu_j)^2}{\lambda_i + \mu_j} |\langle u_i, v_j \rangle|^k \quad \text{for any } k \in \mathbb{N},$$

where $\lambda_1, \ldots, \lambda_n$ and $\mu_1, \ldots, \mu_n$ are eigenvalues of adjacency matrices of $G, G'$, respectively, and $u_1, \ldots, u_n$ and $v_1, \ldots, v_n$ are the corresponding eigenvectors.

|       | $G_1$   | $G_2$   | $G_3$   | $G_4$   | $G_5$   |
|-------|---------|---------|---------|---------|---------|
| $G_1$ | 0       | 2.5325  | 2.8009  | 4.4449  | 11.3354 |
| $G_2$ | 2.5325  | 0       | 1.5889  | 4.2998  | 11.3355 |
| $G_3$ | 2.8009  | 1.5889  | 0       | 4.2473  | 11.3356 |
| $G_4$ | 4.4449  | 4.2998  | 4.2473  | 0       | 11.3363 |
| $G_5$ | 11.3354 | 11.3355 | 11.3356 | 11.3363 | 0       |

Table 2: To scale values of $\|\mathcal{M}_2(G_i) - \mathcal{M}_2(G_j)\|_2$ we alternatively use the value of $\log(\|\mathcal{M}_2(G_i) - \mathcal{M}_2(G_j)\|_2 + 1)$. Note that if $d(\cdot, \cdot)$ is a distance function, then so is $\psi(d(\cdot, \cdot))$ for $\psi(x) = \log(x+1)$.

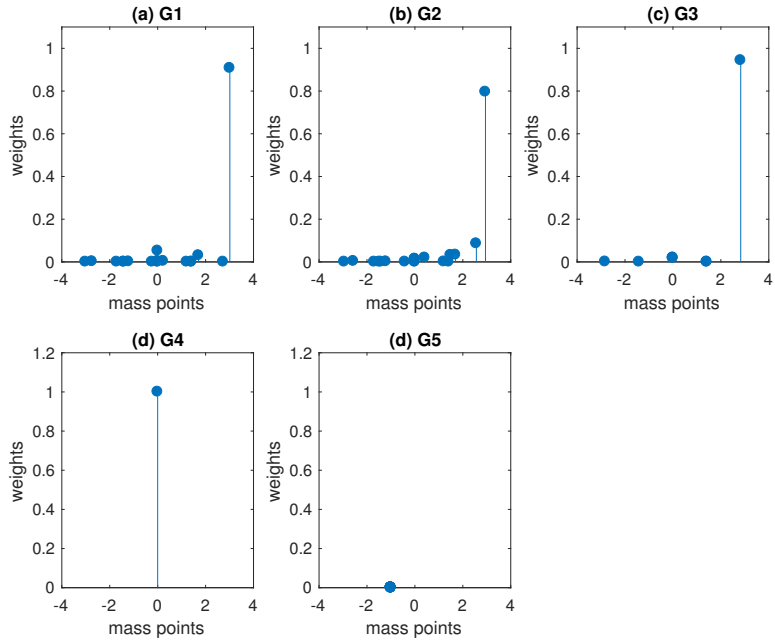

Figure 8: Spectral distributions for graphs $G_1, G_2, G_3, G_4, G_5$.

However, $d(G_1, G_4) = d(G_2, G_4) = d(G_3, G_4) = d(G_5, G_4) = 18$, which means that the graphs $G_1, G_2, G_3, G_5$ are not distinguishable from $G_4$. Table 2 shows that our proposed measure overcomes such drawbacks. As shown in Fig. 8 the spectral distributions for the graphs $G_1, G_2, G_3, G_4, G_5$ are all different. It shows that the distance between two graphs among them follows dissimilarities between the distributions.

|     |       | •  • |  •——•  | •  •  • | △ with • | ¦ • | •—•—• |
|-----|-------|------|--------|---------|----------|-----|-------|
|     |       | (plot) | (plot) | (plot) | (plot) | (plot) | (plot) |
|     |       | $2K_1$ | $K_2$ | $3K_1$ | $K_3$ | $\bar{P}_3$ | $P3$ |
| $2K1$ |     | 0      | 2.8284 | 0      | 20.9762 | 1.8856 | 6.7659 |
| $K2$  |     | 2.8284 | 0      | 2.8284 | 18.7617 | 0.9428 | 4.2164 |
| $3K1$ |     | 0      | 2.8284 | 0      | 20.9762 | 1.8856 | 6.7659 |
| $K3$  |     | 20.9762 | 18.7617 | 20.9762 | 0 | 19.4822 | 14.6211 |
| $\bar{P}_3$ |     | 1.8856 | 0.9428 | 1.8856 | 19.4822 | 0 | 5.0332 |
| $P_3$ |     | 6.7659 | 4.2164 | 6.7659 | 14.6211 | 5.0332 | 0 |

Table 3: Dissimilarities between two graphs among non-isomorphic simple graphs with 2 or 3 vertices.

There are two non-isomorphic simple graphs with 2 vertices and four non-isomorphic simple graphs with 3 vertices. The moment matrix distances $\|M_2(G) - M_2(\tilde{G})\|_2$ between two graphs $G$ and $\tilde{G}$ among them are shown in Table 3. The spectral distributions of their adjacency matrices for each graphs are shown as well. There are 11 non-isomorphic simple graphs with 4 vertices. The spectral distributions of them are shown as well in Table 4. Table 5 gives their distances.

Denote $J$ as the all-1's matrix and $I$ is the identity.

**Example 4.1.** (1) The complete graph $K_n$ has an adjacency matrix equal to $A = J - I$. Note the eigenvalues of $K_n$ are $n-1$ with multiplicity 1 and $-1$ with multiplicity $n-1$. The corresponding eigenvector of $-1$ is $e$. Thus, the spectral measure of $K_n$ in $\varphi_e$ is $\mu(K_n) = \delta_{n-1}$.

(2) Let $G$ be a d-regular graph and $A$ be a adjacency matrix of $G$. It is each to check that $d$ is an eigenvalue of $A$ and the corresponding eigenvector is $e$. Thus, the spectral measure of $G$ in $\varphi_e$ is $\mu(G) = \delta_d$.

(3) Let $K_{mn}$ be the complete bipartite graph and $A$ be a adjacency matrix of $G$. Note that the eigenvalues of $A$ consists of $\sqrt{mn}, -\sqrt{mn}, 0$. Then the moments $m_k$ is $1, \frac{2mm}{m+n}, \frac{mn^2}{m+n}, \frac{m^2n}{m+n}$.

| Name | Graph | Spectral distribution | Name | Graph | Spectral distribution |
|---|---|---|---|---|---|
| $4K_1 = \overline{K_4}$ | | | $K_4 = W_3$ | | |
| co-diamond | | | diamond | | |
| co-paw | | | paw=3-pan | | |
| $2K_2 = \overline{C_4}$ | | | $C_4 = K_{2,2}$ | | |
| claw $= K_{1,3}$ | | | co-claw | | |
| $P_4$ | | | | | |

Table 4: Spectral distributions of 11 non-isomorphic simple graphs with 4 vertices.

## 5. Complexity and Parallelism

Our proposed method has two steps. Consider the moment matrix with degree $n$ whose size is $(n+1) \times (n+1)$. The first step is to obtain the moment matrix $\mathcal{M}_n$ whose entries consist of the moment sequence $\{m_k\}_{k=0}^{2n}$. In the second step, we use Frobenius distance between positive definite matrices to compute the distance between two moment matrices. In the following, we will show the time complexity, space complexity, and parallelism of each step and those of the overall algorithm.

|            | $4K_1$  | $K_4$   | co-diamond | diamond | co-paw  | paw     | $2K_2$  | $C_4$   | claw    | co-claw | $P_4$   |
|------------|---------|---------|------------|---------|---------|---------|---------|---------|---------|---------|---------|
| $4K_1$     | 0       | 90.9945 | 1.4142     | 49.8999 | 5.0744  | 26.2726 | 2.8284  | 20.9762 | 12.3693 | 15.7321 | 9.8742  |
| $K_4$      | 90.9945 | 0       | 90.0777    | 41.4970 | 86.6646 | 65.3854 | 89.1740 | 70.8802 | 79.4292 | 75.8650 | 82.0945 |
| co-diamond | 1.4142  | 90.0777 | 0          | 48.9081 | 3.7749  | 25.1942 | 1.4142  | 19.8494 | 11.1803 | 14.6116 | 8.6313  |
| diamond    | 49.8999 | 41.4970 | 48.9081    | 0       | 45.3900 | 23.9322 | 47.9375 | 29.4279 | 38.0657 | 34.4891 | 40.7247 |
| co-paw     | 5.0744  | 86.6646 | 3.7749     | 45.3900 | 0       | 21.5754 | 2.5981  | 16.1787 | 7.4666  | 10.9659 | 4.8734  |
| paw        | 26.2726 | 65.3854 | 25.1942    | 23.9322 | 21.5754 | 0       | 24.1506 | 5.5000  | 14.1863 | 10.6184 | 16.8300 |
| $2K_2$     | 2.8284  | 89.1740 | 1.4142     | 47.9375 | 2.5981  | 24.1506 | 0       | 18.7617 | 10.0499 | 13.5462 | 7.4498  |
| $C_4$      | 20.9762 | 70.8802 | 19.8494    | 29.4279 | 16.1787 | 5.5000  | 18.7617 | 0       | 8.7750  | 5.2440  | 11.3798 |
| claw       | 12.3693 | 79.4292 | 11.1803    | 38.0657 | 7.4666  | 14.1863 | 10.0499 | 8.7750  | 0       | 3.6742  | 2.7386  |
| co-claw    | 15.7321 | 75.8650 | 14.6116    | 34.4891 | 10.9659 | 10.6184 | 13.5462 | 5.2440  | 3.6742  | 0       | 6.2450  |
| $P_4$      | 9.8742  | 82.0945 | 8.6313     | 40.7247 | 4.8734  | 16.8300 | 7.4498  | 11.3798 | 2.7386  | 6.2450  | 0       |

Table 5: The moment matrix distances $\|\mathcal{M}_2(G) - \mathcal{M}_2(\tilde{G})\|_2$ between two graphs $G$ and $\tilde{G}$ among them are shown.

5.1. Complexity

We consider comparing two graphs $G_1$ and $G_2$. Let $|V_1|, |E_1|$ and $|V_2|, |E_2|$ denote the number of nodes and edges of graph $G_1$ and $G_2$ respectively. Let $|E| = max(|E_1|, |E_2|)$ and $|V| = max(|V_1|, |V_2|)$. The first step of the algorithm can be computed in $\mathcal{O}(n|E|)$ time and $\mathcal{O}(|E|)$ space using sparse matrix-vector multiplication. The second step mainly involves eigenvalue decomposition, which can be computed in $\mathcal{O}(n^3)$ time and $\mathcal{O}(n^2)$ space. The time complexity of the total algorithm is $\mathcal{O}(n|E| + n^3)$ and space complexity is $\mathcal{O}(|E| + n^2)$. However, $n$ is relatively small, say 4 or 5, in practical problems because most of the information about a distribution is contained in the first few moments[18, 27, 22]. Thus the time complexity and space complexity of proposed method are both $\mathcal{O}(|E|)$.

5.2. Parallelism

As we discussed before, the first step is sparse matrix-vector multiplication. This operation can be completely paralleled on CPU or GPU. As $n$ is small, the second step takes much less time than the first step. As a result, our algorithm can be paralleled efficiently.

6. Experiments

6.1. Clustering Networks

We first demonstrate the efficacy of our method and other methods utilizing moment via clustering random networks. Specifically, we generates four sets of Erdős-Rényi random graphs[17]. The parameters are $\Theta_1 = \{|V| = 1000, |E| = 10000, \rho = 0.1\}$, $\Theta_2 = \{|V| = 1000, |E| = 20000, \rho = 0.1\}$, $\Theta_3 = \{|V| = 1000, |E| = 10000, \rho = 0.9\}$, $\Theta_3 = \{|V| = 1000, |E| = 20000, \rho = 0.9\}$, in which $|V|$ denotes the number of nodes, $|E|$ denotes the number of edges, $\rho$ denotes the rewiring probability, i.e. randomness. For example, $\rho = 0$ the network is regular graph while $\rho = 1$ the network is completely random network. For each parameter setting, we generate 25 networks and label the networks according to their parameter settings.

|          | Proposed Method | Cov | NCLM | EIGS | GK3 | GK4 |
|----------|-----------------|-----|------|------|-----|-----|
| ACCURACY | 1               | 1   | 1    | 0.76 | 0.5 | 0.5 |

Table 6: Accuracy for our proposed method and covariance method in random network clustering.

The benchmark algorithms are as following:

- **Cov** [38]: Covariance method computes the computes the covariance matrix of the vector $[\frac{A^i e}{|A^i e|}]_{i=1}^n$, in which $A$ is the adjacency matrix and $e$ is the vector of all ones. Then Bhattacharya similarity between the corresponding covariance matrix is computed as the distance between two graphs. According to their paper, we take the size of moment matrix $n = 4, 5, 6$ and choose the best one as the benchmark.

- **NCLM** [32]: NCLM first computes the log moment sequence vector $[\log(\frac{\text{tr}(A^i)}{n^i})]_{i=2}^7$ and uses the Euclidean distance between two moment vectors as the distance between corresponding networks.

- **GK** [36]: Graphlet kernel computes the distance between graphs by counting subgraphs with $k$ nodes. Here we use $k = 3$ and $k = 4$.

- **EIGS-10**: The eigenvalues of the adjacency matrix contains much information about the graph and are graph invariant. As a result, we take the biggest 10 eigenvalues for each graph. Then we employ Euclidean distance between the eigenvalues of corresponding adjacency matrices as the distance between two graphs.

For NCLM, EIGS, Covariance and our proposed method, we first compute the distance matrix $D$, in which $D_{ij}$ is the distance between the $ith$ network and $jth$ network. Then we construct the kernel $\mathcal{K} = \exp(-D)$. Finally we apply kernel k-means algorithm to get the clustering result. The clustering performance is shown in Table. I.

We see that methods that involves moment, i.e. our proposed method, Covariance, and NCLM perform best. GK3 and GK4 has trouble in separating the networks with the same number of nodes and same number of edges but of different randomness. EIGS can not distinguish between the parameter setting $\Theta_3$ and $\Theta_4$. This demonstrates that the methods based on moment are able to capture the feature of edge distribution and randomness of the network.

|  | HEP Vs CM | HEP Vs ASTRO | ASTRO Vs CM | Full |
| :---: | :---: | :---: | :---: | :---: |
| Proposed | **0.991** | **0.913** | **0.904** | **0.905** |
| EIGS-10 | 0.981 | 0.879 | 0.861 | 0.820 |
| NCLM | 0.982 | 0.850 | 0.865 | 0.804 |
| Covariance | 0.976 | 0.857 | 0.861 | 0.819 |
| Covariance with SVM | 0.987 | 0.889 | 0.887 | 0.849 |

Table 7: Accuracy for Proposed method and other benchmark methods in collaboration network classification. Best results marked in bold.

*6.2. Classifying Networks*

We apply our method to classify networks. We follow the system setting of [38]. Specifically, we classify one's research area using the information of the graph structure of one's collaboration network. Because researchers in one area usually tightly connected with researchers in that area compared to other areas, it is possible to determine to which area a researcher belongs considering one's collaboration networks. This information can be used for recommendations such as job recommendations and citation recommendations.

Three datasets from [34] are used: high energy physics collaboration network(HEP), condensed matter collaboration network(CM), and astro physics collaboration network(ASTRO). In the network, an undirected edge from $u$ to $v$ means that the author $u$ and the author $v$ are co-authored. We use the method from [38] to generate subgraphs and obtain 415 subgraphs for CM and 1000 subgraphs for HEP and ASTRO respectively. Then we label each sub-graph according to the dataset which it belongs to. The tasks are classifications between each two datasets and among three datasets. For each task, we first split the dataset into 10 folds of the same size. We then combine 9 of the folds as the training set, the left 1 fold as the test set. We repeat this 10 times to compute the average accuracy.

In the classification tasks, we use $k$-nearest-neighbor(KNN) classifier. We set the size of moment matrix $n$ from 2 to 7 and $k$ in KNN from 1 to 10 and choose the best one. The first three benchmark algorithms are Covariance, NCLM, and Top-10 eigenvalues(EIGS-10). In addition, we add the state-of-art method in collaboration network classification, Covariance with SVM[38], which employs SVM as the classifier, as the last benchmark algorithm. The performance of our method and the benchmark algorithm is shown in Table 7.

| $|V|$ | $|E|$ | Proposed | Cov | EIGS-10 | NCLM | GK3 |
|---|---|---|---|---|---|---|
| 2000 | 2000000 | **7.31** | 7.32 | 18.92 | 39.92 | 99.63 |
| 5000 | 1000000 | **1.38** | 1.48 | 85.88 | 533 | 797 |
| 10000 | 2000000 | **3.9** | 4.7 | 353.5 | 27340 | 1757 |
| 50000 | 15000000 | **50** | 68 | 11687 | N/A | N/A |

Table 8: Running time for computing pairwise distance among 100 random networks(in seconds). Fastest method marked in bold.

From the table, we see that with KNN classifier, Covariance, EIGS, and NCLM have similar performance in each task. We also notice that Covariance with SVM performs better than Covariance with KNN. This shows that SVM classifier is more suitable to Covariance method. On top of that, our proposed method not only outperforms various of benchmarks with KNN classifier, but also performs better than Covariance with SVM, the state-of-art method in collaboration classification task in every classification task. This demonstrates the effectiveness of proposed method. This also shows that a few moments can provide enough information for collaboration classification. Besides, proposed method has a significant improvement over the state-of-art method in three collaboration network classification task. This shows proposed method is suitable to classification tasks for sophisticated networks.

6.3. Time Comparison

In this section, we show the efficiency of our algorithm by comparing the running time of proposed method and other methods via a set of experiments. Specifically, in each experiment, we generate 100 Erdős-Rényi random graphs with the same number of nodes and edges. Then we employ proposed method and other methods to get pairwise distances among all possible pairs. For each method, we run 10 times and take the average running time. The number of nodes, number of edges, and the time consumed by different methods are shown in Table 8. Here, we use $4 \times 4$ moment matrix in proposed method, $4 \times 4$ covariance matrix in Covariance and 6 moments in NCLM. All of these experiments are done in MATLAB on the server with an Intel Xeon 2.80 Ghz CPU and 64 GB RAM.

As shown in the table, the time cost of proposed method is cheaper than all the comparing methods. For example, it can compute pairwise distances of 100 random networks with 500000 nodes and 15000000 edges in 50 seconds, which has $1.36\times$ speed up to Covariance method and $233\times$ speed up to EIGS-10. Besides, from the table, proposed method is almost linear in terms of the number of edges. This demonstrates proposed method is scalable to massive networks.

## 7. Conclusion

We considered the adjacency matrix of a graph as a real random variable and proposed a new similarity measure for graphs with a distance between corresponding moment matrices of their spectral distributions. Our proposed method demonstrated state-of-art results in collaboration network classification and turned out to be scalable to large-scale graphs.